\documentclass{ifacconf}

\usepackage{graphicx}      
\usepackage{natbib}        

\usepackage{xcolor}
\usepackage{mathtools}
\renewcommand{\vec}[1]{\mathbf{#1}}
\usepackage{epsfig} 
\usepackage{mathptmx} 
\usepackage{times} 
\usepackage{amsmath} 
\usepackage{amssymb}  
\usepackage{leftidx}
\newtheorem{theorem}{Theorem}
\newtheorem{remark}{Remark}
\newtheorem{property}{\textbf{Property}}
\usepackage{color}
\DeclareSymbolFont{newfont}{OML}{cmm}{m}{it}
\DeclareMathSymbol{\Varrho}{3}{newfont}{37}

\DeclareMathOperator{\diag}{diag}

\usepackage{algorithm,algorithmic}
\usepackage{bm}

\begin{document}
\begin{frontmatter}

\title{Adaptive Control of Euler-Lagrange Systems under Time-varying State Constraints without a Priori Bounded Uncertainty \thanksref{footnoteinfo}} 


\thanks[footnoteinfo]{Corresponding author. \\
This work has been partially funded by the European Unions Horizon 2020 Research and Innovation Programme AERO-TRAIN under the Grant Agreement No. 953454. \\
\textbf{\textcopyright~ the authors. Accepted by IFAC for publication
under a Creative Commons Licence CC-BY-NC-ND}}

\author{Viswa Narayanan Sankaranarayanan$^1$*},
\author{Sumeet Satpute$^1$},
\author{Spandan Roy$^2$} and
\author{George Nikolakopoulos$^1$}

\address{$^1$ Robotics and Artificial Intelligence Group, Department of Computer Science, Electrical and Space Engineering, Lule\aa \,\, University of Technology, Sweden (e-mail: (vissan, sumsat, geonik)@ltu.se)}
\address{$^2$ Robotics Research Center, International Institute of Information Technology, Hyderabad, India (spandan.roy@iiit.ac.in)}

\begin{abstract}
In this article, a novel adaptive controller is designed for Euler-Lagrangian systems under predefined time-varying state constraints. The proposed controller could achieve this objective without a priori knowledge of system parameters and, crucially, of state-dependent uncertainties. The closed-loop stability is verified using the Lyapunov method, while the overall efficacy of the proposed scheme is verified using a simulated robotic arm compared to the state of the art.
\end{abstract}

\begin{keyword}
Adaptive control, barrier Lyapunov function, Euler-Lagrangian system
\end{keyword}

\end{frontmatter}

\section{Introduction} \label{sec:intro}
Euler-Lagrangian (EL) models of system dynamics represent a plethora of real-world electro-mechanical systems, ranging from simple engineering applications to cutting-edge engineering fields, such as space technology and robotics (cf. \cite{spong2008robot}). Additionally, they pose exciting challenges to control theory owing to practical issues, such as modeling uncertainties, external disturbances, etc., and therefore, over the years, various robust and adaptive control techniques (cf. \cite{utkin2013adaptive, lu2017antiswing, zhou2019adaptive, sankaranarayanan2020aerial, berberich2022combining, laghrouche2021barrier}) have been explored to address the problems of modeling uncertainties and external disturbances. 

On the other hand, safety is a crucial aspect in various applications, such as aerial and surgical robotics, space engineering, and robot-human interactions, where breaching a certain level of control accuracy may cause severe damage to both the robot and the environment. Safety is usually ensured by imposing state constraints. A few popular model-based methods in this direction are the model predictive control (\cite{mayne2006robust}), the zeroing Lyapunov functions (\cite{ames2016control}), and the invariant set applications (\cite{blanchini1999set}). However, the high computational burden and the dependency on a priori knowledge of the system model limited applications of these controllers for complex systems with significant modeling uncertainty. 


Subsequently, a low complexity method based on Barrier Lyapunov Function (BLF) was conceptualized to handle state and output constraints (\cite{tee2009barrier}). While initial works on BLF-based controllers relied on precise knowledge of the system model, recent works reported various controllers (cf. \cite{ganguly2021efficient, xu2021tangent, cruz2022non}) and adaptive (cf. \cite{liu2017barrier, shao2021adaptive, pang2019adaptive, yang2020state, smaeilzadeh2018finite, obeid2018barrier, laghrouche2021barrier}) to overcome limited knowledge on system model and disturbances. 

Nevertheless, the aforementioned BLF-based controllers only consider fixed-value constraints. This fact limits the operational range of the controllers by restricting the initial condition within a narrow range, which is undesirable in many practical applications (cf. the discussions later in Remark 2). Therefore, BLF-based control techniques have adopted time-varying constraint functions (cf. \cite{liu2020finite, wang2021barrier, li2021adaptive, ding2021adaptive, liu2019adaptive, fuentes2020adaptive, sachan2020synthesis, sankaranarayanan2022robustifying, sankaranarayanan2022robust}). However, these works either require precise system model knowledge (cf. \cite{liu2020finite, wang2021barrier, li2021adaptive, ding2021adaptive, sankaranarayanan2022robustifying, sankaranarayanan2022robust}) or can only tackle a priori bounded uncertainty (cf. \cite{liu2019adaptive, fuentes2020adaptive, sachan2020synthesis}). Crucially, for EL systems, uncertainty functions are inherently state-dependent, i.e., they cannot be considered to be bounded a priori to the stability analysis: in fact, it has been shown in (\cite{roy2020new}) that adaptive control laws developed considering a priori bounded uncertainty can even lead to instability.

In view of the above discussions and to the best of the authors' knowledge, an adaptive control solution for EL systems, under time-varying state constraints without a priori bounded uncertainty, is still missing in the literature. In this direction, this research brings out the following contributions:
\begin{itemize}
    \item The proposed adaptive BLF-based controller can negotiate time-varying state constraints.
    \item The proposed adaptive law can handle unknown state-dependent uncertainties and hence, suitable for EL systems.
\end{itemize}
The closed-loop system stability is studied analytically. Further, improved performance of the proposed controller is noted compared to state-of-the-art in a simulated environment using a pick-and-place robot.


The rest of the article is organized as follows. Section \ref{sec:system_model} introduces the EL system dynamics; Section \ref{sec:con_des} establishes the design of the proposed controller, and Section \ref{sec:sta_ana} describes the proposed adaptive controller and its closed-loop stability analysis, respectively; Section \ref{sec:sim_res} presents the simulated results and the related discussions, while Section \ref{sec:concl} provides concluding remarks.

\textbf{Notations:} The notation $\mathbf{I}$ denoted the identity matrix of appropriate dimensions; $||\mathbf{a}||, \lambda_{\min}(\mathbf{a})$ denote Euclidean norm and minimum eigenvalue of the vector $\mathbf{a}$, respectively and a diagonal matrix with $a,b,...,n$ as its diagonal elements is denoted by, diag$\lbrace a,b,...,n \rbrace$.


\section{System Dynamics and Problem Formulation} \label{sec:system_model}
Consider the following standard EL systems dynamics (\cite{spong2008robot}):
\begin{align}
& \mathbf M( {\mathbf q}(t))\ddot{ \mathbf q}(t)+\mathbf C({\mathbf q}(t),\dot{\mathbf q}(t))\dot{\mathbf q}(t)+ \nonumber \\
& \qquad \qquad \mathbf G(\mathbf q(t))+ \mathbf F(\dot{\mathbf q}(t))+\mathbf{ d}(t)=\boldsymbol \tau(t), \label{eq:sys_dyn}
\end{align} 
where $\mathbf{q}, \dot{\mathbf q} \in\mathbb{R}^{n}$ are the system states; $\mathbf{M(q)}\in\mathbb{R}^{n\times n}$ is the mass/inertia matrix; $\mathbf C(\mathbf q,\dot{\mathbf q})\in\mathbb{R}^{n\times n}$ denotes the Coriolis/centripetal terms; $\mathbf G(\mathbf q)\in\mathbb{R}^{n}$ denotes the gravity vector; $\mathbf F(\dot{\mathbf q})\in\mathbb{R}^{n}$ represents the vector of damping and friction forces; $\mathbf{ d}(t) \in \mathbb{R}^n$ denotes bounded external disturbance and $\boldsymbol \tau \in \mathbb{R}^n$ is the generalized control input.

The EL-dynamics \eqref{eq:sys_dyn} presents the following system properties (cf. \cite{spong2008robot}), which are useful for control design and analysis: 
\begin{property} \label{pro:mass_inertia}
$\mathbf{M}(\mathbf{q})$ is uniformly positive definite $\forall \mathbf{q}$, and there exists scalars $\underline{m}, \overline{m} ~ \in \mathbb{R}^{+}$ such that,
\begin{align*}
    0 < \underline{m}\mathbf{I} \leq \mathbf{M}({\mathbf{q}}) \leq \overline{m}\mathbf{I}. 
\end{align*}
\end{property}
\begin{property} \label{pro:bounds}
There exists scalars $\overline c, \overline{g}, \overline f, \overline{d} \in \mathbb{R}^{+}$, such that $||\mathbf{C(q},\dot{\mathbf q})|| \leq \overline c ||\dot{\mathbf q}||$, $||\mathbf G(\mathbf q)|| \leq \overline g$, $||\mathbf{F}(\dot{\mathbf q})|| \leq \overline f ||\dot{\mathbf q}||$ and $||\mathbf d(t)|| \leq \overline{d}$, $\forall \mathbf q,\dot{\mathbf q}$, $\forall t \geq 0$.
\end{property}
\begin{property} \label{pro:skew_sym}
The matrix $(\dot{\mathbf{M}}(\mathbf{q}) -2{\mathbf{C}( \mathbf{q},\dot{\mathbf{q}})})$ is skew symmetric, i.e., for any non-zero vector $\mathbf{y}$, we have $\mathbf{y}^T(\dot{\mathbf{M}}(\mathbf{q})-2\mathbf{C(q},\dot{\mathbf{q}}))\mathbf{y}=0$.
\end{property}
\begin{assum}[Knowledge of the system dynamics] \label{re:state_dep}
The knowledge of dynamic terms $\mathbf{M, ~ C, ~ G, ~ F, ~ d}$ and of their bounds (cf. Properties 1-2) is unavailable for the control design.
\end{assum}
\begin{remark}
  Assumption 1 implies that the control objective (outlined subsequently) is to achieved without any a priori knowledge of the dynamics terms and the disturbances.
\end{remark}
Further, let us define the following standard assumption on the desired trajectory (cf. \cite{tee2009control}).
\begin{assum} \label{as:des_traj}
The desired trajectory $\mathbf{q}_{d}(t)$ is designed to be smooth and bounded.
\end{assum}

The objective to track a desired trajectory by keeping the states within some designated constraints can be equivalently posed as the control objective to keep tracking errors within some designated bounds. Accordingly, let us define tracking error $\mathbf{e} (t) \triangleq \mathbf{q}(t) - \mathbf{q}_{d}(t) = \lbrace e_1(t), e_2(t), \cdots, e_n(t) \rbrace$ and time-varying error constraints $\mathbf{b}(t)$ given by
\begin{subequations} \label{eq:bounds}
\begin{align}
    \mathbf{b}(t) &= [ b_1(t) , b_2(t), \cdots,  b_n(t)], \\
    b_i(t) &= (\rho_{0i} - \rho_{ssi})\exp({-\omega_{i} t}) + \rho_{ssi},~ \rho_{0i}> e_i (0), \label{eq:p} 
\end{align}
\end{subequations}
where the user-defined scalars $\rho_{0i}, \rho_{ssi}$ denote the initial and final values of the constraint $b_i$, $i \in \lbrace 1,2,..,n \rbrace$ respectively, such that $\rho_{0i} \geq \rho_{ssi}$; $\omega_{i}$ are design scalars orchestrating the rate of convergence of the constraints, which is useful to tune the transients.

\begin{remark}
    Compared to a conventional (adaptive) control strategy, BLF-based design can contain the tracking error within predefined bounds. Hence, a time-varying bound, whose initial value is larger than the initial error and exponentially converges to a smaller steady-state value (as in \eqref{eq:bounds}), is well-suited for practical applications. Further, the parameter $\omega_{i}$ allows the user to design the convergence rate based on the mechanical constraints.
\end{remark}

The control problem is defined as follows:

\textbf{Control Problem:} Design an adaptive controller for the EL system \eqref{eq:sys_dyn} under the Properties \ref{pro:mass_inertia} - \ref{pro:skew_sym} and Assumptions 1-2 that ensures tracking error remains within a predefined accuracy without any a priori knowledge of the system dynamics and the bounds of the uncertainties (cf. Remark 1).

\section{Proposed Control Solution} \label{sec:con_des}
Let us define an error variable
\begin{align}
    \mathbf{s} &= \mathbf{\dot{e}} +  
                \boldsymbol{\Phi} \boldsymbol{\psi} \label{eq:s}
\end{align}
where $\boldsymbol{\Phi}$ is a user-defined positive definite matrix and
\begin{subequations}
\begin{align}
    \boldsymbol{\psi} &=  \boldsymbol{\zeta} \mathbf{e}, \label{eq:psi_p} \\
    \boldsymbol{\zeta} &= \diag \left \lbrace \frac{1}{b^2_1 - {e}^2_1},  \frac{1}{b^2_2 - {e}^2_2}, ...,  \frac{1}{b^2_n - {e}^2_n} \right \rbrace. \label{eq:zeta_p}
\end{align}
\end{subequations}
The time derivative of \eqref{eq:s} yields
\begin{align}
\dot{\mathbf{s}} &=    \ddot{\mathbf{q}} - \ddot{\mathbf{q}}_{d}  +   \boldsymbol{\Phi}\boldsymbol{\zeta}\mathbf{\dot{e}} +  \boldsymbol{\Phi}\boldsymbol{\dot{\zeta}} \mathbf{e}  \label{eq:s_dot_1} 
\end{align}
where
\begin{align*}
    \dot{\boldsymbol{\zeta}} &= \diag \left \lbrace \dot{\zeta}_{1}, \dot{\zeta}_{2}, ..., \dot{\zeta}_{n}  \right \rbrace, \\ 
    \dot{\zeta}_{i} &= -\frac{2 b_i \dot{b}_i - 2 e_i \dot{e}_i}{(b^2_i - e^2_i)^2}, \\
    \dot{b}_i &= -\omega_{i} (\rho_{0i} - \rho_{ssi})\exp(-\omega_{i} t). \nonumber
\end{align*}
The term $\dot{\boldsymbol{\zeta}}$ in (\ref{eq:s_dot_1}) can be rewritten as
\begin{align}
    \dot{\boldsymbol{\zeta}} &= \boldsymbol{\zeta}^2 (\mathbf{r}_1 + \mathbf{r}_2 ) \label{eq:zeta_p_dot_2} \\
    \mathbf{r}_1 &=  -2 \diag \left \lbrace
    b_1 \dot{b}_1, ..., 
    b_n \dot{b}_n \right \rbrace, \nonumber \\
    \mathbf{r}_2 &= 2 \diag \left \lbrace {e}_1 \dot{e}_1,...,{e}_n \dot{e}_n    \right \rbrace.
\end{align}
Multiplying both sides of \eqref{eq:s_dot_1} by $\mathbf{M}$, and substituting \eqref{eq:sys_dyn}, \eqref{eq:zeta_p_dot_2} lead to
\begin{align}
    \mathbf{M}\dot{\mathbf{s}} &=  \mathbf{M}  \left( \ddot{\mathbf{q}} - \ddot{\mathbf{q}}_{d}  + 
    \boldsymbol{\Phi} \left (\boldsymbol{\zeta} \mathbf{\dot{e}}   +  \boldsymbol{\zeta}^2 (\mathbf{r}_1 + \mathbf{r}_2) \mathbf{e} \right)  \right) \nonumber  \\
    &= \boldsymbol{\tau} - \mathbf{C}\mathbf{s} + \boldsymbol{\varphi}, \label{eq:ms_dot_3)} \\
    \text{where }\quad \boldsymbol{\varphi} &= - \left(\mathbf{C} \dot{\mathbf{q}} + \mathbf{G} + \mathbf{F} + \mathbf{d} + \mathbf{M}\ddot{\mathbf{q}}_{d} \right) + \mathbf{C} \left(\mathbf{\dot{e}} +  \boldsymbol{\Phi} \boldsymbol{\psi}  \right) \nonumber \\
    & \quad + \mathbf{M} \boldsymbol{\Phi} \left (\boldsymbol{\zeta} \mathbf{\dot{e}}   +  \boldsymbol{\zeta}^2 (\mathbf{r}_1 + \mathbf{r}_2) \mathbf{e} \right) \label{eq:phi}
\end{align}
is the overall uncertainty in the system.
Using the relationship \eqref{eq:phi} and Properties \ref{pro:mass_inertia}-\ref{pro:bounds}, $\boldsymbol{\varphi}$ can be upper bounded as 
\begin{align}
    ||\boldsymbol{\varphi}|| \leq & \underbrace{(\overline{c} ||\mathbf{\dot{q}}_d|| + \overline{f} ||\mathbf{\dot{q}}_d|| + \overline{g} + \overline{d} + \overline{m} ||\mathbf{\ddot{q}}_d||)  }_{K_0} \underbrace{(1)}_{\chi_0} + \underbrace{\overline{f}}_{K_1} \underbrace{||\boldsymbol{\xi}||}_{\chi_1} \nonumber \\
    & \quad + \underbrace{(\overline{m} + \overline{c}) ||\boldsymbol{\Phi}|| }_{K_2} \underbrace{||\boldsymbol{\zeta}|| ||\boldsymbol{\xi}||}_{\chi_2} \nonumber \\
    & \quad + \underbrace{2 \overline{m}||\mathbf{r}_1||}_{K_3} \underbrace{||\boldsymbol{\zeta}||^2 ||\boldsymbol{\xi}||}_{\chi_3} + \underbrace{2 \overline{m}}_{K_4} \underbrace{||\boldsymbol{\zeta}||^2 ||\boldsymbol{\xi}||^3}_{\chi_4} \nonumber \\
    \leq & \sum_{j=0}^{4} K_j \chi_j,
\end{align}
where $K_j$ are unknown scalars. 
\begin{remark}
Note that the upper bound of the uncertainty function $\boldsymbol{\varphi}$ has an explicit dependency on the states via $\chi_j$ and hence, cannot be assumed to be bounded by a constant a priori to the stability analysis. 
\end{remark}

The control law is proposed as
\begin{align}
    \boldsymbol{\tau} &= - \boldsymbol{\Lambda} \mathbf{s} - \sum_{j=0}^{4} \hat{K}_j \chi_j (\mathbf{s}/||\mathbf{s}||), \label{eq:control_law}
\end{align}
where $\boldsymbol{\Lambda}$ is a user-defined positive definite gain and $\widehat{K}_j$ are the estimates of $K_j$ evaluated by the following adaptive laws 
\begin{align}
    \dot{\hat{K}}_j = ||\mathbf{s}||\chi_j - \eta_j \hat{K}_j, \label{eq:adap_law}
\end{align}
with $\eta_j$ being user-defined positive design scalars.

\section{Closed-Loop Stability analysis} \label{sec:sta_ana}
\begin{theorem}
    Under Properties \ref{pro:mass_inertia} - \ref{pro:skew_sym}, Assumptions 1-2, and using the control law (\ref{eq:control_law}) and the adaptive law (\ref{eq:adap_law}), the trajectory of the closed-loop system (\ref{eq:ms_dot_3)}) remains Uniformly Ultimately Bounded (UUB), and the error trajectories $\mathbf{e}, \dot{\mathbf{e}}$ remain within the bounds defined in (\ref{eq:bounds}) for all $t \geq 0$.
\end{theorem}
\textit{Proof:} The solutions for the adaptive gains (\ref{eq:adap_law}) can be derived to be
\begin{align}
    \hat{K}_j = & \underbrace{\exp(- \eta_j t) \hat{K}_j(0)}_{\geq 0}
                + \underbrace{\int_{0}^{t} \exp(- \eta_j(t-\vartheta)) (||s||\chi_j)d \vartheta}_{\geq 0} \nonumber \\
    \implies \hat{K}_j \geq & 0, ~ j \in \lbrace 0,1,...,4 \rbrace, ~ \forall t \geq 0. \label{eq:Kgeq0}
\end{align}

The closed-loop stability of the system is analyzed using the following Lyapunov-like function,
\begin{align}
    V =& \frac{1}{2} \mathbf{s}^T \mathbf{M \,s} + \frac{1}{2}\sum_{j=0}^{4}\left(\hat{K}_j - K_j \right)^2 . \label{eq:V} 
\end{align}
Using the dynamics defined in (\ref{eq:sys_dyn}), the control law (\ref{eq:s})-(\ref{eq:control_law}), the time derivative of $V$ can be expressed as
\begin{align}
    \dot{V} =& \mathbf{s}^T\mathbf{M}\dot{\mathbf{s}} + \frac{1}{2}\mathbf{s}^T \dot{\mathbf{M}} \mathbf{s}  + \sum_{j=0}^{4}\dot{\hat{K}}_j\left(\hat{K}_j - K_j \right)  \\
              =& \mathbf{s}^T \left(-\boldsymbol{\Lambda}\mathbf{s} - \sum_{j=0}^{4} \hat{K}_j \chi_j (\mathbf{s}/||\mathbf{s}||) - \mathbf{C}\mathbf{s} + \boldsymbol{\varphi} \right) \nonumber \\
              & \quad + \frac{1}{2} \mathbf{s}^T \dot{\mathbf{M}} \mathbf{s} + \sum_{j=0}^{4}\dot{\hat{K}}_j\left(\hat{K}_j - K_j \right) . \label{eq:v_dot}
\end{align}
From Properties \ref{pro:mass_inertia}, \ref{pro:skew_sym} and further simplification of \eqref{eq:v_dot}, we can upper bound $\dot{V}$ as
\begin{align}
    \dot{V} \leq & -{\lambda_{\min}(\boldsymbol{\Lambda})||\mathbf{s}||^2} -  \sum_{j=0}^4 \left 
    \lbrace (||\mathbf{s}|| \chi_j - \dot{\hat{K}}_j)(\hat{K}_j - K_j) \right \rbrace . \label{eq:dot_V_01}
\end{align}

From the adaptive law (\ref{eq:adap_law}), we can simplify (\ref{eq:dot_V_01}) as
\begin{align}
    \dot{V} \leq & -{\lambda_{\min}(\boldsymbol{\Lambda})||\mathbf{s}||^2} - \sum_{j=0}^4
     \eta_j\hat{K}_j(\hat{K}_j - K_j) ,\nonumber \\
     \leq & -{\lambda_{\min}(\boldsymbol{\Lambda})||\mathbf{s}||^2} - \sum_{j=0}^4
     (\eta_j \hat{K}^2_j - \eta_j \hat{K}_j K_j), \nonumber \\
     \leq & -{\lambda_{\min}(\boldsymbol{\Lambda})||\mathbf{s}||^2}  - \sum_{j=0}^4 \frac{\eta_j}{2} \left(\left(\hat{K}_j - K_j \right)^2 - K^2_j \right) . \label{eq:dot_V_02}
\end{align}
From \ref{eq:V}, we have
\begin{align}
    V \leq & \frac{\bar{m}}{2}||\mathbf{s}||^2 + \sum_{j=0}^4 \frac{1}{2}(\hat{K}_j - K_j)^2 . \label{eq:V_upper}
\end{align}
Using \eqref{eq:V_upper}, and defining a scalar $\kappa$, such that $0 < \kappa < \Varrho$, $\dot{V}$ from \eqref{eq:dot_V_02} can be simplified as
\begin{align}
    \dot{V} \leq & - \Varrho V + \sum_{j=0}^4 \frac{\eta_j}{2} K^2_j, \\
            = & - \Varrho V - (\Varrho - \kappa) V + \sum_{j=0}^4 \frac{\eta_j}{2} K^2_j, \label{eq:dot_V_eq}
\end{align}
where $\Varrho \triangleq \frac{\min \lbrace \lambda_{\min} (\boldsymbol{\Lambda}), \eta_j \rbrace}{\max \lbrace \bar{m}/2, 1/2 \rbrace}$. Defining a scalar, $\bar{\mathcal{B}} \triangleq \frac{\sum_{j=0}^4 \eta_j K^2_j}{2(\Varrho - \kappa)}$, it can be seen that $\dot{V} < - \kappa V(t)$ when $V>\bar{\mathcal{B}}$, so that
\begin{align}
    V \leq & \max \lbrace V(0), \bar{\mathcal{B}} \rbrace, ~ \forall t>0
\end{align}
implying that the closed-loop system is UUB (cf. \cite{khalil2002nonlinear} for definition) implying that $\mathbf{s}$ and $K_j$ remain bounded. {This implies that} the tracking error trajectories $e_i$ and their time-derivatives $\dot{e}_i$ with $i=\lbrace x,y,\theta \rbrace$ do not violate the state constraints $b_i$ and $b_{di}$ respectively for all $t \geq 0$.
\begin{remark}[Continuity in control law]
In practice, to avoid control chatter, the term $\mathbf{s/||s||}$ can be replaced by the saturation function (cf. \cite{roy2020towards}). This does not change the overall UUB result, albeit minor changes in the stability analysis.
\end{remark}

\section{Simulation Results and Analysis} \label{sec:sim_res}

\subsection{Simulation Scenario} 
\begin{figure*}
    \centering
    \includegraphics[width=0.95\textwidth]{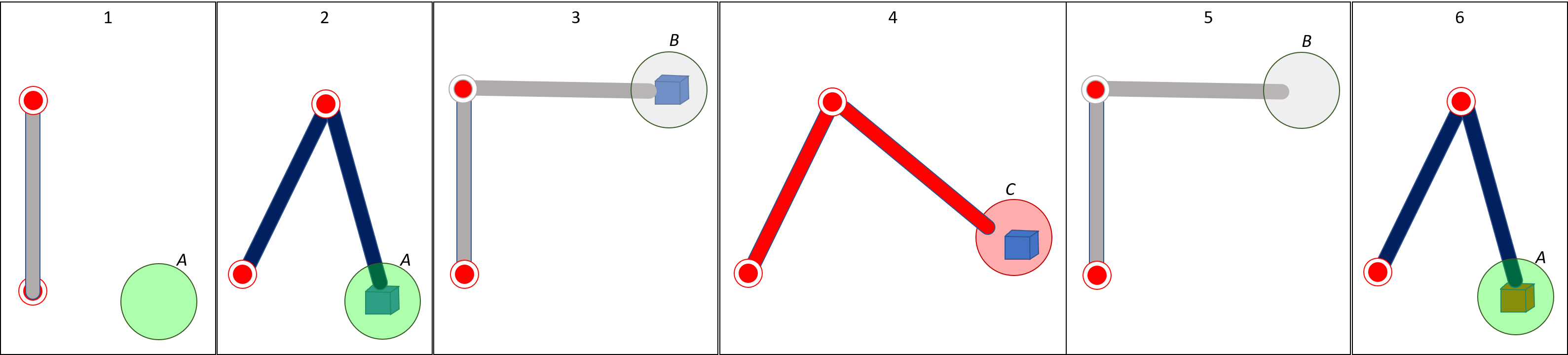}
    \caption{The sequence of operation of the manipulator. (1) From any initial point, the manipulator settles at the initial position. (2) From the initial position, the end-effector is moved to region A, where it picks the payloads from conveyor 1. (3) To avoid collision, the end-effector is moved to an intermediate location in region B. (4) The end-effector is moved to region C, where it can drop the payloads in converyor 2. (5) The manipulator returns after dropping the payload via region B. (6) The end-effector picks up a new payload.}
    \label{fig:arm_seq}
\end{figure*}

The performance of the proposed controller is verified in the MATLAB-SIMULINK environment using a 2-link robotic manipulator, which conducts a pick-and-place operation by transferring payloads from one conveyor to another. The robot loads the payload from region A and carries it through region B to unload it in region C (cf. Fig. \ref{fig:arm_seq}).

\begin{figure}[!h]
    \centering
    \includegraphics[width=0.45\textwidth, trim={0 0 0 0}]{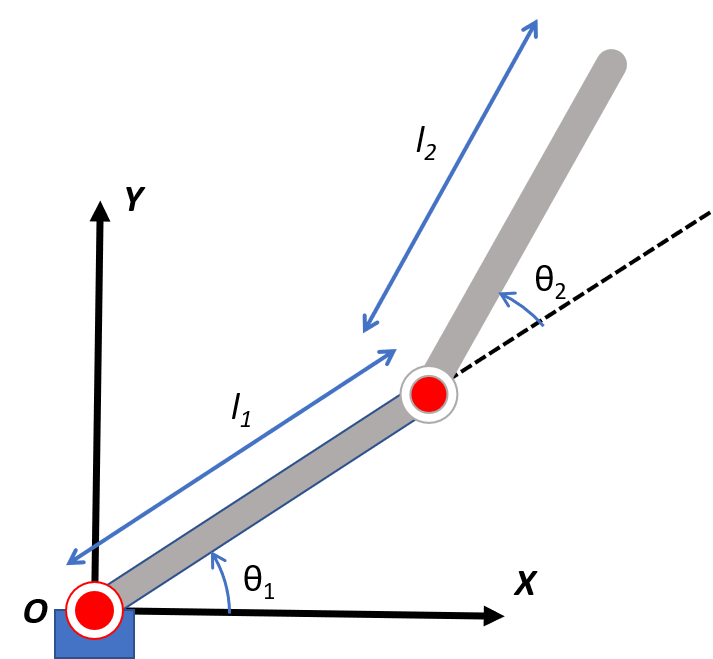}
    \caption{A schematic of the manipulator with its coordinate frame.}
    \label{fig:arm_frame}
\end{figure}
Further, Fig. \ref{fig:arm_frame} shows a schematic of the manipulator's links and coordinate frame, where the lengths of the links are denoted by $l_1, l_2$, and the joint angles are given by $\theta_1, \theta_2$. The dynamics of the manipulator follows (\ref{eq:sys_dyn}) with
\begin{align*}
    \mathbf{M} = \begin{bmatrix}
    m_{11} & m_{12} \\ m_{21} & m_{22}
    \end{bmatrix}, & ~~
    \mathbf{C} = \begin{bmatrix}
    c_{211} \dot{\theta}_2 & c_{121} \dot{\theta}_1 + c_{221} \dot{\theta}_2 \\ c_{112} \dot{\theta}_1 & 0
    \end{bmatrix},\\
    \mathbf{G} = \begin{bmatrix}
    G_1 \\ G_2
    \end{bmatrix}, & ~~ 
    \mathbf{F} = \begin{bmatrix}
    \gamma_1 \\ \gamma_2
    \end{bmatrix} \mathbf{\dot{q}}, ~~
    \mathbf{d} = \begin{bmatrix}
        d_1 \\ d_2
    \end{bmatrix},
\end{align*}
where
\begin{align*}
    m_{11} &= m_1 r^2_1 + m_2 \left(l^2_1 + r^2_1 + 2 l_1 r_1 \cos(\theta_2) \right) + I_1 + I_2, \\
    m_{12} &= m_2 \left ( l_1 r_2 \cos(\theta_2) + r^2_2 \right ) + I_2, \\ m_{21} &= m_{21}, \quad m_{22} = m_2 r^2_2 + I_2, \\
    c_{112} &= c_{121} = c_{211} = c_{221} = -m_2 l_1 r_2 \sin(\theta_2), \\
    G_{1} &= m_1 g \cos(\theta_1) + m_2 g \left (l_1 \cos(\theta_1) + r_2 \cos(\theta_1 + \theta_2) \right), \\
    G_{2} &= m_2 g r_2 \cos(\theta_1 + \theta_2), \\
    I_1 &= m_1 \left (3 r^2 + l^2_1  \right ), \quad I_2 = m_2 \left (3 r^2 + l^2_2  \right ), \\
    d_1 &= 2 \sin(0.1 t), \quad d_2 = 2 \cos(0.1 t)
\end{align*}
where $\gamma_1 = \gamma_2 = 0.1$ are the coefficients of frictions for the joints, $r = 0.05$m is the radius of the links (considering the links to be regular cylinders); $l_1 = l_2 = 1$m, $r_1 = l_1/2, r_2 = l_2/2$ are the positions of the center of masses of the two links along their lengths; $m_1 = m_2 = 1 $kg are the masses of the links, and $g = 9.81 \text{m/s}^2$ is the acceleration due to gravity. The bounds are set to be $\rho_{01} = 120^0$, $\rho_{02} = 185^0$ $\rho_{ss1} = \rho_{ss2} = 3^0$ , $\omega_{1} = \omega_{2} = 0.1, \boldsymbol{\Phi} = \diag \lbrace 1.2, 1.2 \rbrace, \hat{K}_{j}(0) = 0.1 \forall j \in \lbrace 0,1,...,4 \rbrace$, and $\boldsymbol{\Lambda} = \diag \lbrace 4, 4 \rbrace $.

\begin{figure}[!h]
    \centering
    \includegraphics[width=0.48\textwidth]{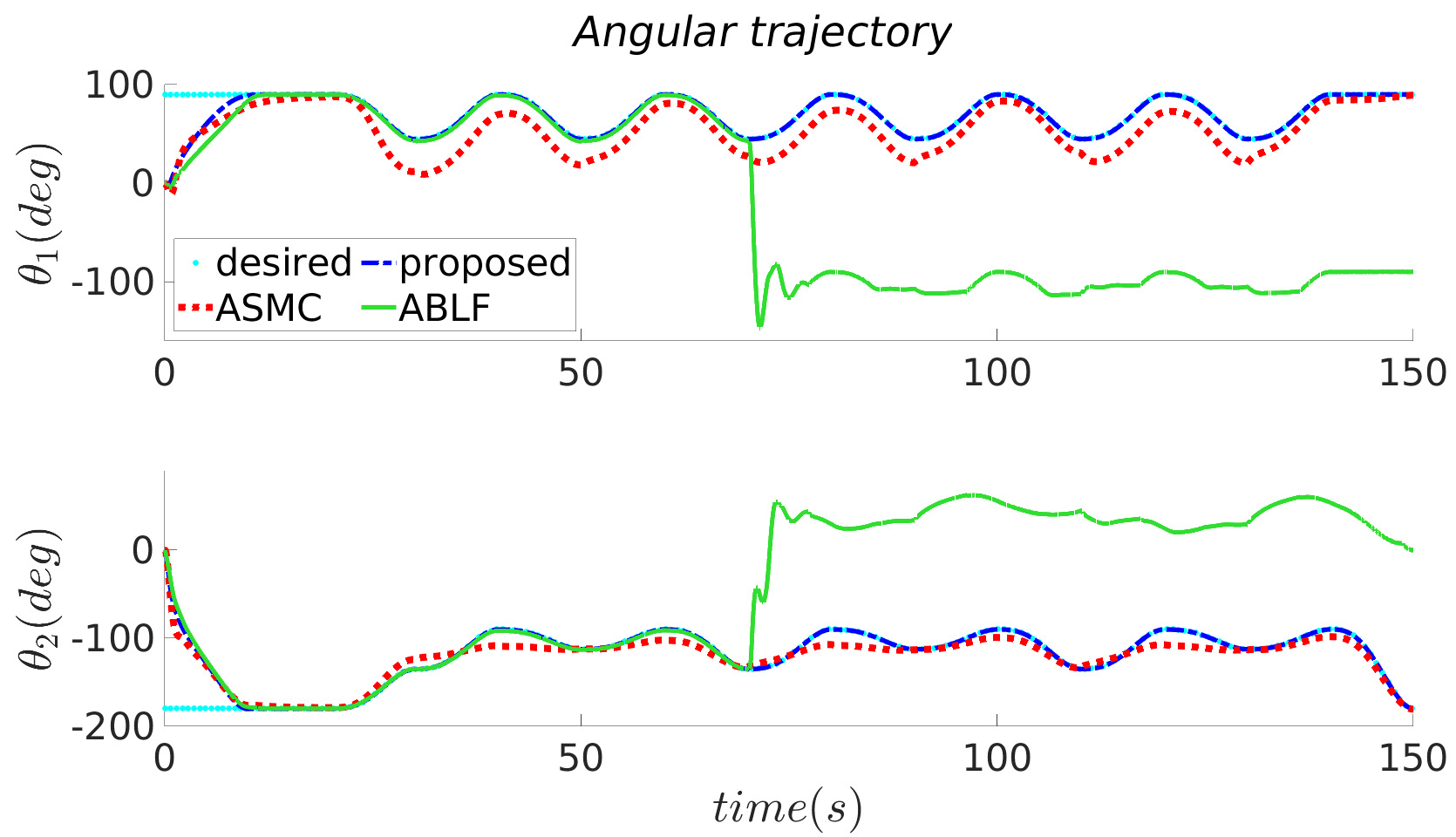}
    \caption{Trajectory of the angles $\theta_{1},\theta_{2}$ obtained by different controllers along with the desired trajectory.}
    \label{fig:ang_traj}
\end{figure}
The sequence of operation is explained below (cf. Fig. \ref{fig:arm_seq}):

\begin{enumerate}
    \item From its initial position, the unloaded manipulator is set to the starting orientation by commanding an initial orientation of $\theta_1(0) = 90^0$, $\theta_2(0) =-180^0$. 
    \item From this orientation, the end effector is moved to region A by providing desired angles of $\theta_{1d} = 45^0, \theta_{2d} = -135^0$. The first payload (mass$=0.5 $kg) from conveyor 1 is attached to the arm at this position ($t=30$s).
    \item The payload is moved to region B to avoid collision while moving to conveyor 2. The desired angles are set to be $\theta_{1d} = 90^0, \theta_{2d} = -90^0$.
    \item The end effector is moved to region C, where is unloads the payload on conveyor 2 with desired angles $\theta_{1d} = 45^0, \theta_{2d} = -135^0$.
    \item Then the unloaded arm is moved to reach region B before moving to region A for collecting the next payload.
    \item The process is repeated with payloads of masses $1 $kg, $1.5 $kg at $t=70$s and $t=110$s, respectively.
\end{enumerate}

To properly judge the effectiveness, the performance of the proposed controller is compared with Adaptive Sliding Mode Controller (ASMC) as in (\cite{shao2021adaptive}) and with adaptive time-varying BLF (ABLF) controller as in (\cite{liu2020finite}).

\subsection{Results}
\begin{figure}[!h]
    \centering
    \includegraphics[width=0.48\textwidth] {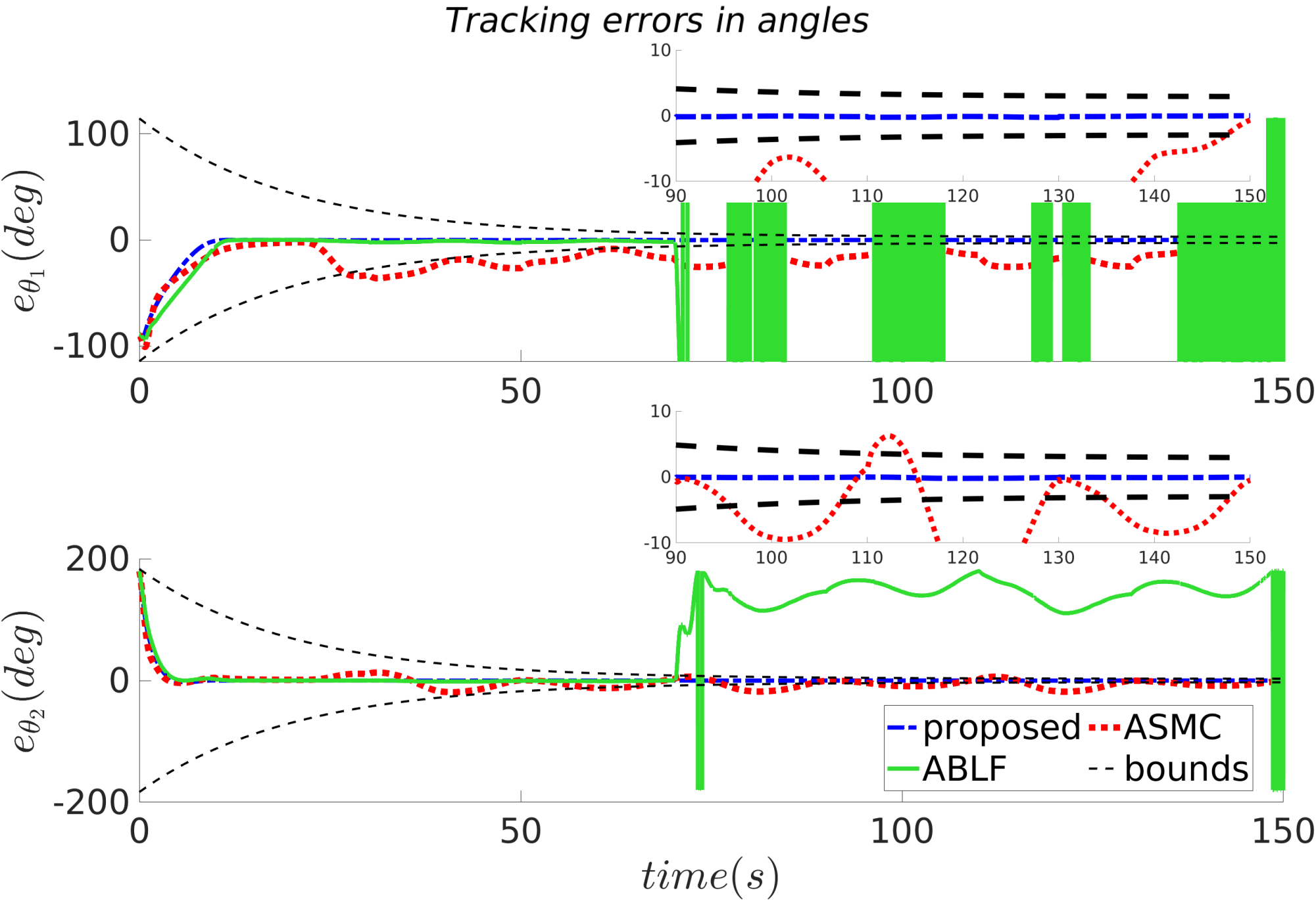}
    \caption{Errors in angles $\theta_{1},\theta_{2}$. The proposed controller ensures that the errors stay within bounds.}
    \label{fig:ang_err}
\end{figure}
\begin{figure}[!h]
    \centering
    \includegraphics[width=0.48\textwidth]{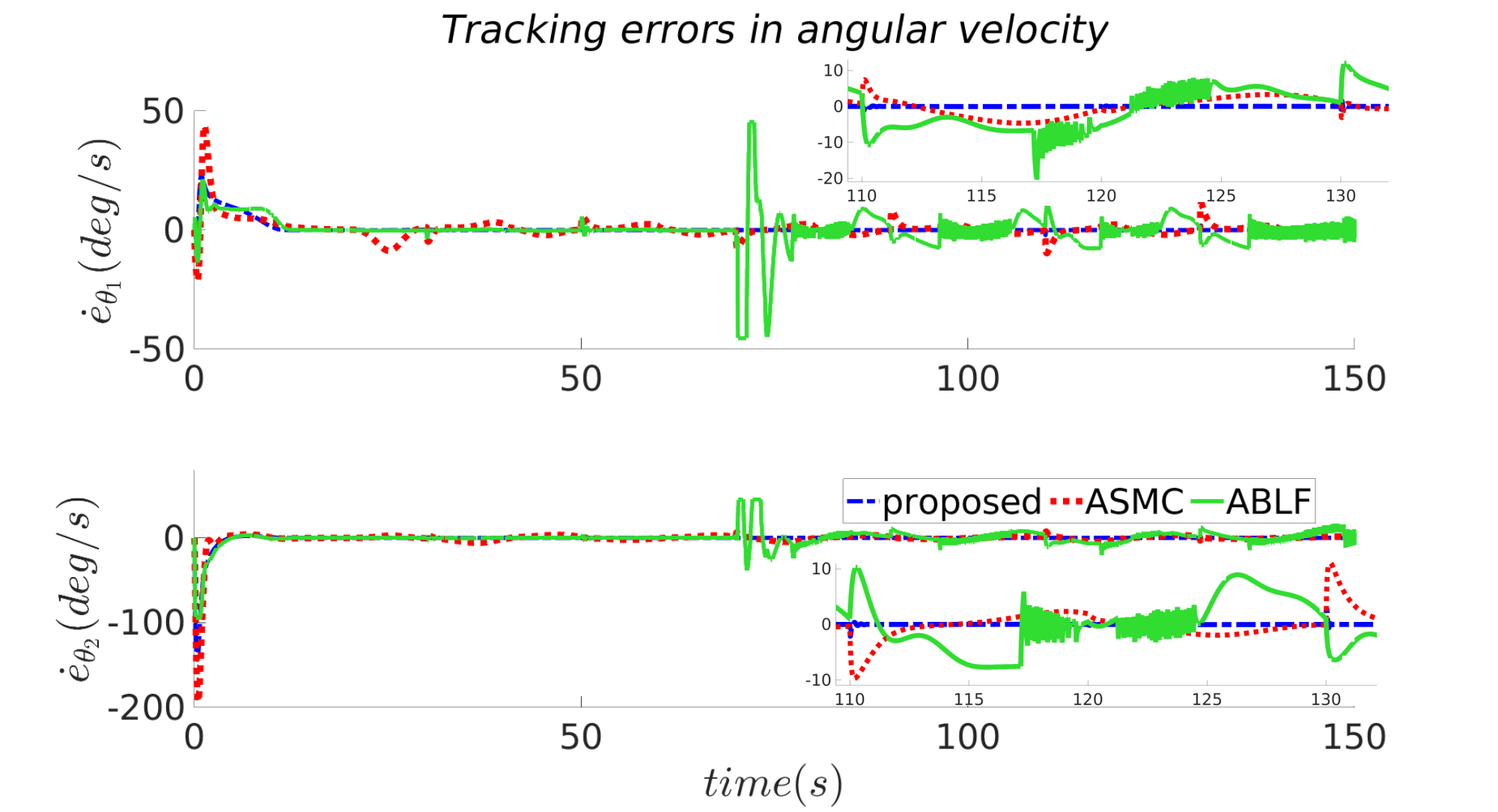}
    \caption{Errors in angular velocities $\dot{\theta}_{1},\dot{\theta}_{2}$. The proposed controller ensures that the errors stay within the bounds.}
    \label{fig:ang_vel_err}
\end{figure}
\begin{figure}[!h]
    \centering
    \includegraphics[width=0.48\textwidth]{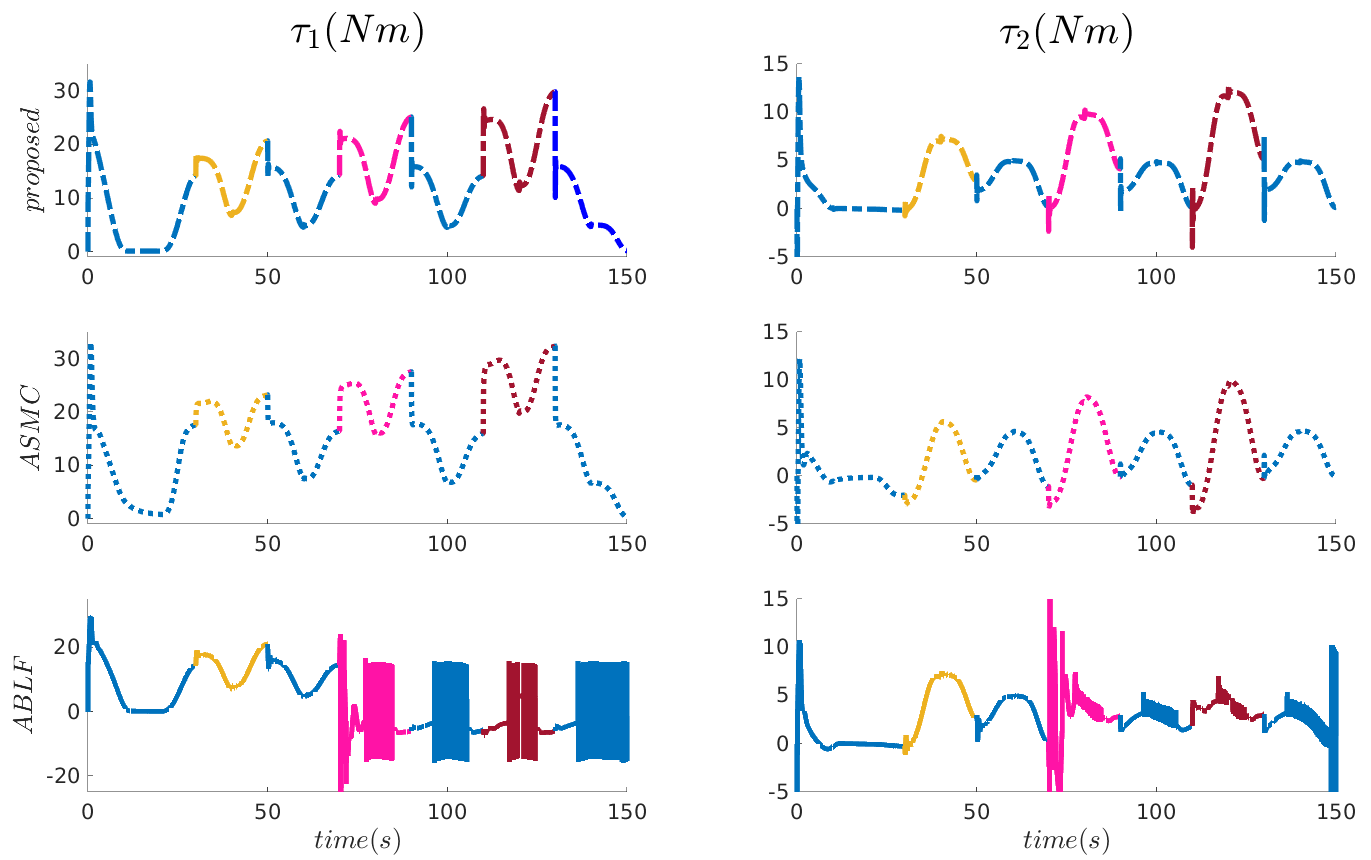}
    \caption{Torque inputs $\tau_{1}, \tau_{2}$ generated by the controllers for the respective joints. The blue lines represent the unloaded period; the yellow lines represent the period with the first payload; the pink lines represent the period with the second payload, and the brown lines represent the period with the third payload.}
    \label{fig:control_input}
\end{figure}


The performances of the various controllers are demonstrated via Figs. \ref{fig:ang_traj}-\ref{fig:control_input}.
The violation of error constraints (or bounds) is quite evident from the error plots for ASMC as it is not designed to handle state constraints. On the other hand, ABLF closely tracks the desired trajectory within the state constraints till the first payload is added; however, from $t \geq 70$s, when payloads of higher masses are added, undesirable high frequency oscillations can be observed (cf. Fig. \ref{fig:ang_err}). This happens because the change in mass due to payload causes state-dependent uncertainty via the dynamics terms $\mathbf{M} \ddot{\mathbf q}$ and $\mathbf{C} \dot{\mathbf q}$. ABLF is not designed to tackle such uncertainty, and hence, when error increases toward the boundary value of constraints, high frequency instability occurs. Note that, such oscillations did not arise during the addition of the first payload of lesser weight. This shows that stability under ABLF depends on the amount of uncertainty induced, and in case of unknown uncertainty, stability cannot be guaranteed. A similar trend is observed in Fig. \ref{fig:ang_vel_err}.

Whereas the proposed controller maintains the error within the bounds during the addition and removal of all three payloads (cf. the magnified snippets, in which the ABLF error is removed).  
The root-mean-squared (RMS) values and peak values of the errors are tabulated in Table. \ref{tb:per_com}. For a fair comparison, the peak values are calculated using the error values after $t=20$s (when all the error trajectories approach zero for the first time). Further, for the ABLF controller, the values from $t=20-69$s (just before the second payload is added) are used for computing the metrics. The tabulated data shows the remarkable quality of the proposed controller in disturbance rejection and adaptation to
changes in dynamics compared to ASMC and ABLF.

\begin{table}[!h]
\caption{{Performance comparison of the controllers}}
\label{tb:per_com}
		\centering
{
{	\begin{tabular}[b]{c c c c c}
		\hline
		\hline
		& \multicolumn{2}{c}{RMS error (deg)} & \multicolumn{2}{c}{RMS velocity error (deg/s)}  \\ \cline{1-5}
		  & $\theta_1$  & $\theta_2$ & $\dot{\theta}_1$ & $\dot{\theta}_2$ \\
		 \hline
		ABLF \cite{} & 2.296 & 1.458 & 0.463 & 0.363 \\
		 \hline
		ASMC \cite{} & 19.450 & 9.316 & 2.217 & 2.172 \\
		\hline
		proposed & 0.182 & 0.114 & 0.093 & 0.070 \\
		\hline 
		\% Improvement & 92.07 & 92.18 & 79.91 & 80.72 \\
		\hline
		\hline
		& \multicolumn{2}{c}{Peak error (deg)} & \multicolumn{2}{c}{Peak velocity error (deg/s)}  \\ \cline{1-5}
		  & $\theta_1$  & $\theta_2$ & $\dot{\theta}_1$ & $\dot{\theta}_2$ \\
		 \hline
		ABLF \cite{} & 2.677 & 1.792 & 5.783 & 3.970 \\
		 \hline
		ASMC \cite{} & 36.279 & 19.313 & 11.166 & 7.422 \\
		\hline
		proposed & 0.445 & 0.285 & 2.884 & 2.123 \\
		\hline
\end{tabular}}}
\end{table}

\section{Conclusions} \label{sec:concl}
An adaptive controller ensuring state constraints is designed without the knowledge of both the system dynamics and the upper bound of the uncertainties. The incorporated state constraints are time-varying, which removes the dependence of the steady-state constraints on the initial conditions of the state errors. The adaptive gains are designed to overcome unknown state-dependent uncertainties in the system. The closed-loop stability is analyzed analytically. The proposed controller's performance is verified and compared to a conventional adaptive controller and a BLF-based adaptive controller, using a 2D-robotic manipulator, performing a pick-and-place operation in a Simulink environment.

\bibliography{final_submit}

\begin{thebibliography}{33}
\providecommand{\natexlab}[1]{#1}
\providecommand{\url}[1]{\texttt{#1}}
\providecommand{\urlprefix}{URL }
\expandafter\ifx\csname urlstyle\endcsname\relax
  \providecommand{\doi}[1]{doi:\discretionary{}{}{}#1}\else
  \providecommand{\doi}{doi:\discretionary{}{}{}\begingroup \urlstyle{rm}\Url}\fi

\bibitem[{Ames et~al.(2016)Ames, Xu, Grizzle, and Tabuada}]{ames2016control}
Ames, A.D., Xu, X., Grizzle, J.W., and Tabuada, P. (2016).
\newblock Control barrier function based quadratic programs for safety critical systems.
\newblock \emph{IEEE Transactions on Automatic Control}, 62(8), 3861--3876.

\bibitem[{Berberich et~al.(2022)Berberich, Scherer, and Allg{\"o}wer}]{berberich2022combining}
Berberich, J., Scherer, C.W., and Allg{\"o}wer, F. (2022).
\newblock Combining prior knowledge and data for robust controller design.
\newblock \emph{IEEE Transactions on Automatic Control}.

\bibitem[{Blanchini(1999)}]{blanchini1999set}
Blanchini, F. (1999).
\newblock Set invariance in control.
\newblock \emph{Automatica}, 35(11), 1747--1767.

\bibitem[{Cruz-Ortiz et~al.(2022)Cruz-Ortiz, Chairez, and Poznyak}]{cruz2022non}
Cruz-Ortiz, D., Chairez, I., and Poznyak, A. (2022).
\newblock Non-singular terminal sliding-mode control for a manipulator robot using a barrier lyapunov function.
\newblock \emph{ISA transactions}, 121, 268--283.

\bibitem[{Ding et~al.(2021)Ding, Zhang, Mei, and Park}]{ding2021adaptive}
Ding, S., Zhang, B., Mei, K., and Park, J.H. (2021).
\newblock Adaptive fuzzy sosm controller design with output constraints.
\newblock \emph{IEEE Transactions on Fuzzy Systems}.

\bibitem[{Fuentes-Aguilar and Chairez(2020)}]{fuentes2020adaptive}
Fuentes-Aguilar, R.Q. and Chairez, I. (2020).
\newblock Adaptive tracking control of state constraint systems based on differential neural networks: a barrier lyapunov function approach.
\newblock \emph{IEEE transactions on neural networks and learning systems}, 31(12), 5390--5401.

\bibitem[{Ganguly et~al.(2021)Ganguly, Sankaranarayanan, Suraj, Yadav, and Roy}]{ganguly2021efficient}
Ganguly, S., Sankaranarayanan, V.N., Suraj, B., Yadav, R.D., and Roy, S. (2021).
\newblock Efficient manoeuvring of quadrotor under constrained space and predefined accuracy.
\newblock In \emph{2021 IEEE/RSJ International Conference on Intelligent Robots and Systems (IROS)}, 6352--6357. IEEE.

\bibitem[{Khalil(2002)}]{khalil2002nonlinear}
Khalil, H.K. (2002).
\newblock \emph{Nonlinear systems}, volume~3.
\newblock Prentice hall Upper Saddle River, NJ.

\bibitem[{Laghrouche et~al.(2021)Laghrouche, Harmouche, Chitour, Obeid, and Fridman}]{laghrouche2021barrier}
Laghrouche, S., Harmouche, M., Chitour, Y., Obeid, H., and Fridman, L.M. (2021).
\newblock Barrier function-based adaptive higher order sliding mode controllers.
\newblock \emph{Automatica}, 123, 109355.

\bibitem[{Li(2021)}]{li2021adaptive}
Li, G. (2021).
\newblock Adaptive sliding mode control for a class of manipulator systems with output constraint.
\newblock \emph{Complexity}, 2021.

\bibitem[{Liu et~al.(2020)Liu, Liu, Wang, Zhou, and Lu}]{liu2020finite}
Liu, C., Liu, X., Wang, H., Zhou, Y., and Lu, S. (2020).
\newblock Finite-time adaptive tracking control for unknown nonlinear systems with a novel barrier lyapunov function.
\newblock \emph{Information Sciences}, 528, 231--245.

\bibitem[{Liu et~al.(2019)Liu, Ma, Liu, Tong, and Chen}]{liu2019adaptive}
Liu, Y.J., Ma, L., Liu, L., Tong, S., and Chen, C.P. (2019).
\newblock Adaptive neural network learning controller design for a class of nonlinear systems with time-varying state constraints.
\newblock \emph{IEEE Transactions on Neural Networks and Learning Systems}, 31(1), 66--75.

\bibitem[{Liu and Tong(2017)}]{liu2017barrier}
Liu, Y.J. and Tong, S. (2017).
\newblock Barrier lyapunov functions for nussbaum gain adaptive control of full state constrained nonlinear systems.
\newblock \emph{Automatica}, 76, 143--152.

\bibitem[{Lu et~al.(2017)Lu, Fang, Sun, and Wang}]{lu2017antiswing}
Lu, B., Fang, Y., Sun, N., and Wang, X. (2017).
\newblock Antiswing control of offshore boom cranes with ship roll disturbances.
\newblock \emph{IEEE Transactions on Control Systems Technology}, 26(2), 740--747.

\bibitem[{Mayne et~al.(2006)Mayne, Rakovi{\'c}, Findeisen, and Allg{\"o}wer}]{mayne2006robust}
Mayne, D.Q., Rakovi{\'c}, S.V., Findeisen, R., and Allg{\"o}wer, F. (2006).
\newblock Robust output feedback model predictive control of constrained linear systems.
\newblock \emph{Automatica}, 42(7), 1217--1222.

\bibitem[{Obeid et~al.(2018)Obeid, Fridman, Laghrouche, and Harmouche}]{obeid2018barrier}
Obeid, H., Fridman, L.M., Laghrouche, S., and Harmouche, M. (2018).
\newblock Barrier function-based adaptive sliding mode control.
\newblock \emph{Automatica}, 93, 540--544.

\bibitem[{Pang et~al.(2019)Pang, Zhang, and Xu}]{pang2019adaptive}
Pang, H., Zhang, X., and Xu, Z. (2019).
\newblock Adaptive backstepping-based tracking control design for nonlinear active suspension system with parameter uncertainties and safety constraints.
\newblock \emph{ISA transactions}, 88, 23--36.

\bibitem[{Roy and Baldi(2020)}]{roy2020towards}
Roy, S. and Baldi, S. (2020).
\newblock Towards structure-independent stabilization for uncertain underactuated euler--lagrange systems.
\newblock \emph{Automatica}, 113, 108775.

\bibitem[{Roy et~al.(2020)Roy, Lee, and Baldi}]{roy2020new}
Roy, S., Lee, J., and Baldi, S. (2020).
\newblock A new adaptive-robust design for time delay control under state-dependent stability condition.
\newblock \emph{IEEE Transactions on Control Systems Technology}, 29(1), 420--427.

\bibitem[{Sachan and Padhi(2020)}]{sachan2020synthesis}
Sachan, K. and Padhi, R. (2020).
\newblock Synthesis of an adaptive state-constrained control for mimo euler-lagrange systems.
\newblock \emph{IFAC-PapersOnLine}, 53(2), 5518--5523.

\bibitem[{Sankaranarayanan et~al.(2020)Sankaranarayanan, Roy, and Baldi}]{sankaranarayanan2020aerial}
Sankaranarayanan, V.N., Roy, S., and Baldi, S. (2020).
\newblock Aerial transportation of unknown payloads: Adaptive path tracking for quadrotors.
\newblock In \emph{2020 IEEE/RSJ International Conference on Intelligent Robots and Systems (IROS)}, 7710--7715. IEEE.

\bibitem[{Sankaranarayanan et~al.(2022{\natexlab{a}})Sankaranarayanan, Yadav, Swayampakula, Ganguly, and Roy}]{sankaranarayanan2022robustifying}
Sankaranarayanan, V.N., Yadav, R.D., Swayampakula, R.K., Ganguly, S., and Roy, S. (2022{\natexlab{a}}).
\newblock Robustifying payload carrying operations for quadrotors under time-varying state constraints and uncertainty.
\newblock \emph{IEEE Robotics and Automation Letters}, 7(2), 4885--4892.

\bibitem[{Sankaranarayanan et~al.(2022{\natexlab{b}})Sankaranarayanan, Banerjee, Satpute, and Nikolakopoulos}]{sankaranarayanan2022robust}
Sankaranarayanan, V.N., Banerjee, A., Satpute, S., and Nikolakopoulos, G. (2022{\natexlab{b}}).
\newblock A robust post-grasping control design for robotic testbed demonstration of space debris disposal.
\newblock \emph{IFAC-PapersOnLine}, 55(38), 198--203.

\bibitem[{Shao et~al.(2021)Shao, Tang, Xu, Wang, and Zheng}]{shao2021adaptive}
Shao, K., Tang, R., Xu, F., Wang, X., and Zheng, J. (2021).
\newblock Adaptive sliding mode control for uncertain euler--lagrange systems with input saturation.
\newblock \emph{Journal of the Franklin Institute}, 358(16), 8356--8376.

\bibitem[{Smaeilzadeh and Golestani(2018)}]{smaeilzadeh2018finite}
Smaeilzadeh, S.M. and Golestani, M. (2018).
\newblock Finite-time fault-tolerant adaptive robust control for a class of uncertain non-linear systems with saturation constraints using integral backstepping approach.
\newblock \emph{IET Control Theory \& Applications}, 12(15), 2109--2117.

\bibitem[{Spong et~al.(2008)Spong, Hutchinson, and Vidyasagar}]{spong2008robot}
Spong, M.W., Hutchinson, S., and Vidyasagar, M. (2008).
\newblock \emph{Robot dynamics and control}.
\newblock John Wiley \& Sons.

\bibitem[{Tee and Ge(2009)}]{tee2009control}
Tee, K.P. and Ge, S.S. (2009).
\newblock Control of nonlinear systems with full state constraint using a barrier lyapunov function.
\newblock In \emph{Proceedings of the 48h IEEE Conference on Decision and Control (CDC) held jointly with 2009 28th Chinese Control Conference}, 8618--8623. IEEE.

\bibitem[{Tee et~al.(2009)Tee, Ge, and Tay}]{tee2009barrier}
Tee, K.P., Ge, S.S., and Tay, E.H. (2009).
\newblock Barrier lyapunov functions for the control of output-constrained nonlinear systems.
\newblock \emph{Automatica}, 45(4), 918--927.

\bibitem[{Utkin and Poznyak(2013)}]{utkin2013adaptive}
Utkin, V.I. and Poznyak, A.S. (2013).
\newblock Adaptive sliding mode control.
\newblock In \emph{Advances in sliding mode control}, 21--53. Springer.

\bibitem[{Wang et~al.(2021)Wang, Liu, He, and Shen}]{wang2021barrier}
Wang, D., Liu, S., He, Y., and Shen, J. (2021).
\newblock Barrier lyapunov function-based adaptive back-stepping control for electronic throttle control system.
\newblock \emph{Mathematics}, 9(4), 326.

\bibitem[{Xu et~al.(2021)Xu, Tang, and Liu}]{xu2021tangent}
Xu, F., Tang, L., and Liu, Y.J. (2021).
\newblock Tangent barrier lyapunov function-based constrained control of flexible manipulator system with actuator failure.
\newblock \emph{International Journal of Robust and Nonlinear Control}, 31(17), 8523--8536.

\bibitem[{Yang et~al.(2020)Yang, Gao, Cui, and Ma}]{yang2020state}
Yang, D., Gao, X., Cui, E., and Ma, Z. (2020).
\newblock State-constraints adaptive backstepping control for active magnetic bearings with parameters nonstationarities and uncertainties.
\newblock \emph{IEEE Transactions on Industrial Electronics}, 68(10), 9822--9831.

\bibitem[{Zhou et~al.(2019)Zhou, Wen, Wang, and Yang}]{zhou2019adaptive}
Zhou, J., Wen, C., Wang, W., and Yang, F. (2019).
\newblock Adaptive backstepping control of nonlinear uncertain systems with quantized states.
\newblock \emph{IEEE Transactions on Automatic Control}, 64(11), 4756--4763.

\end{thebibliography}
\end{document}